\documentclass[11pt,oneside]{article}
\usepackage{graphicx}
\begin{document}
 
\title{{\bf Relationship between the Velocity Ellipsoids of Galactic-Disk
Stars and their Ages and Metallicities}}
\author{{\bf V.~V.~Koval', V.~A.~Marsakov, T.~V.~Borkova}\\
Institute of Physics, Southern Federal University,\\
Rostov-on-Don, Russia\\
e-mail: koval@ip.rsu.ru, marsakov@ip.rsu.ru, borkova@ip.rsu.ru}
\date{accepted \ 2009, Astronomy Reports, Vol. 53 No. 9, P.785-800}
\maketitle

\begin {abstract}
Abstract The dependences of the velocity ellipsoids of F--G stars of the
thin disk of the Galaxy on their ages and metallicities are analyzed 
based on the new version of the Geneva.Copenhagen Catalog. The age 
dependences of the major, middle, and minor axes of the ellipsoids, 
and also of the dispersion of the total residual veltocity, obey power 
laws with indices 0.25, 0.29, 0.32, and 0.27 (with uncertainties $\pm 0.02$).
Due to the presence of thick-disk objects, the analogous indices for all
nearby stars are about a factor of 1.5 larger. Attempts to explain such values
are usually based on modeling relaxation processes in the Galactic 
disk. Elimination of stars in the most numerous moving groups from the 
sample slightly reduces the corresponding indices (0.22, 0.26, 0.27, and
0.24). Limiting the sample to stars within 60\,pc of the Sun, so that the
sample can be considered to be complete, leaves both the velocity  ellipsoids
and their age dependences virtually unchanged. With increasing age, the 
velocity ellipsoid increases in size and becomes appreciably more spherical,
turns toward the direction of the Galactic center, and loses angular momentum.
The shape of the velocity ellipsoid remains far from equilibrium. With increasing
metallicity, the velocity ellipsoid for stars of mixed age increases in size,
displays a weak tendency to become more spherical, and turns toward the direction
of the Galactic center (with these changes occurring substantially more rapidly
in the transition through the metallicity $[Fe/H]\approx -0.25$). Thus, the
ellipsoid changes similarly to the way it does with age; however, with decreasing
metallicity, the rotational velocity about the Galactic center monotonically
increases, rather than decreases (!). Moreover, the power-law indices for
the age dependences of the axes depend on the metallicity, and display a
maximum near $[Fe/H]\approx -0.1$. The age dependences of all the velocity-ellipsoid
parameters for stars with equal metallicity are roughly the same. It is proposed
that the appearance of a metallicity dependence of the velocity ellipsoids
for thin-disk stars is most likely due to the radial migration of stars.
\\


\end {abstract}

\section*{Introduction}

The observed morphological structure of the thin
disk of our Galaxy, like that of any of its subsystems,
is due purely to the collected orbits of the stars
making it up. Therefore, the ''external'' appearance
of the Galaxy can be reconstructed based on the
space velocities of stars, even those located in the
immediate vicinity of the Sun (assuming, of course,
that this location is not distinguished in any way).
Simultaneously, we can attempt to trace the dynamical
evolution of the disk system using stars of various
ages. A classic method for deriving such information
is analysis of the age dependences of the parameters
of the Schwartschild velocity ellipsoids of stars close
to the Sun. First and foremost, it is important to
elucidate how the dispersions of the stellar-velocity
components vary with age. The increases of all the
ellipsoid axes with age can be described well by power
laws of the form $\sigma_{i} \sim t^{\gamma}$, where $\sigma_{i}$ is 
the dispersion of
the corresponding velocity component, $t$ the age of the stars, and 
$\gamma$ the pover-law index. It is usually supposed that these dependences
arise due to relaxation processes in the Galaxy
(see, for example, [1,\,2]). (The alternative explanation
that the elocities of the stars remain undistorted from
birth, so that the age dependence of the velocity dispersion
reflects variations in the dynamical state of
the interstellar medium with time, is thought to be
less likely.) The power-law index provides information
about inhomogeneities of the gravitational potential
of the Galaxy that lead to a continuous growth in
the velocity dispersions of stars that are born, i.\,e.,
to ''heating'' of these stars. In particular, numerical
modeling of heating by stochastic spiral density
waves is able to explain the observed age—velocity-dispersion
dependence over a broad range of power-law
indices $(0.2 < \gamma < 0.7)$. These numerical calculations
can be used to place constraints, including
constraints on parameters of the spiral structure of
the Galaxy~[3]. The velocities may also be perturbed
by a bar in the central region of the Galaxy~[4]. However,
the spiral waves and the bar ''operate'' only in
a thin layer of the Galactic disk, and are unable to
explain the increase in the vertical velocity component.
In this connection, molecular clouds are invoked,
which can explain power-law indices in the
$Z$ direction up to values $\gamma \approx 0.26$~[5], 
whereas higher indices require clusters of dark matter 
from decaying companion galaxies disrupted by the tidal forces of
our Galaxy~[6].

A new era in studies of stellar kinematics arrived
with the publication of the data from the HIPPARCOS
satellite. Thanks to these new precision
radial-velocity and proper-motion measurements, it
has become clear that the velocity distribution for
field stars is not uniform, with a large number of
groups of stars having the same angular momenta
being distinguished. These so-called moving groups
or stellar flows can be arbitrarily separated into two
types according to their origin. The first type of stellar
flow is associated with irregularities in the Galactic
potential. In particular, the Pleiades-Hyades and
Sirius flows can be explained by heating of the disk by
stochastic spiral waves, whose theoretical possibility
was demonstrated in~[3]. Another possible perturber
of the stellar velocity ellipsoid is a bar at the Galactic
center, whose existence is suggested by infrared observations
of stars in that region~[7]. It is likely that
precisely a bar that generates spiral density waves~[8]
led to the formation of the $\xi$-\,Hercules flow (branch) at
the outer Lindblad resonance [9,\,10], which is located
in velocity phase space at the boundary between the
subsystems of the thick and thin disks of the Galaxy.
Another type of flow is associated with the remnants
of disrupted, fairly massive ($4\times 10^{8}M_{\odot}$) companions
to the Milky Way~[11]. Several methods for distinguishing
stars in moving groups have been developed,
which all lead to the identification of essentially the
same stellar flows with the same members. According
to numerous studies in the solar neighborhood,
roughly a third of stars can be identified as members
of some moving group (see, for example, [11,\,12]).
The stellar flows formed by inhomogeneities in the
gravitational potention, i.\,e., by spiral waves and the
bar, make up the majority. All these flows distort
the velocity field of field stars of various ages, hindering
the derivation of the information required to
reconstruct the dynamical evolution of the Galaxy.
After excluding stars in flows, the distributions for
the remaining stars in $U-V$ and $V-log t$ diagrams
become smoother~[13].

The relations between age and velocity dispersion
for nearby field stars have already been investigated
using modern astrometric and spectroscopic measurements.
The power-law indices for various samples
of stars lie in the range from $\gamma\approx 0.34$~[14] to
$\gamma\approx 0.50$ [1,\,2]. Among the most recent attempts
to construct age - velocity-dispersion relations we
note~[15], where it was shown that ''heating'' of all the
velocity components of Galactic-disk stars (with the
mean value $\langle\gamma\rangle \approx 0.35$) has occurred 
over the entire lifetime of this subsystem. This conclusion concerning
the continuous action of relaxation processes is
supported by the results of~[2], which made use of
revised ages and metallicities for the stars (increasing
the power-law index only slightly: 
$\langle\gamma\rangle \approx 0.40$). In
their study of the age dependence of the vertical velocity
component using data from the same catalog~[15],
Seabroke and Gilmor~[16] concluded that a power
law was not required to describe this relation, since
the value of $\sigma_{W}$ becomes saturated after $\sim 4.5$ 
billion years. Unfortunately, the relations considered in all these
studies were constructed for nearby field stars, which
include an appreciable number of thick-disk stars.
Moreover, to obtain correct results for a sample of
thin-disk stars, it is necessary to eliminate stars in
moving groups. Fuller information about the kinematics
of stars is provided by studies of the age dependences
of all the velocity-ellipsoid parameters, not
only the dispersions of the velocity components, as
were considered in the papers noted above. In connection
with the much discussed existence of very old
and simultaneously metal-rich stars, it is also desirable
to trace the variations in the velocity ellipsoids for
thin-disk stars with different metallicities. An analysis
of the velocity ellipsoids for thin-disk stars taking
into account all these factors is the goal of the current
study.

\section*{OBSERVATIONAL DATA.}

The main source of data for the study was the new
version of the Geneva.Copenhagen catalog, which
contains the ages, metallicities, and kinematics for
$\approx14 000$ F-\,G dwarfs brighter than $V\approx 8.5^{m}$~[2].
(These same authors continued to refine the parameters
of the catalog stars~[17], but the distances,
temperatures, and ages of the stars changed appreciably
less than in their first revision, leaving the
features of the age~-- velocity-dispersion relations virtually
unchanged, as the authors themselves noted.
The newest version of the catalog is not yet openly
available.) The metallicities in the catalog were determined
from the $ubvy\beta$ photometric indices corrected
for interstellar reddening. The distances for
approximately 75\,\% of the catalog stars were determined
from the HIPPARCOS trigonometric parallaxes,
with only parallaxes with errors less than 13\,\%
being used. Photometric distances were used for stars
without parallaxes or with parallaxes having large
errors; the limiting accuracy for these is again 13\,\%.
Photometric distances were used primarily for distant
stars, while trigonometric parallaxes were used for the
absolute majority of stars within 60\,pc. The radial velocities
in the catalog were obtained at various times
as part of the CORAVEL project, and proper motions
were taken from the ''Tycho'' catalog. The resulting
claimed accuracy in the stellar velocity components
is $\pm 1.5$~km\,s$^{-1}$.

In the new catalog~[2], as in the previous version~[15], 
the most probable stellar ages are calculated
using the method of~[20] and the theoretical
isochrones [18,\,19], taking into account the uncertainties
in the effective temperatures, absolute magnitudes,
and metallicities. The variability of the speeds
with which stars move along the evolutionary tracks
was automatically taken into account. Holmberg et\,al.~[2] 
found overall good agreement between their
ages and the ages of~[21], which were determined
using the same method and modern Yale isochrones.
Allowing for the variability in the speeds of stars along
their evolutionary tracks had a negative effect on stars
located near turning points, where the isochrones
have complex shapes, since stars evolve comparatively
rapidly in such sections, and the probability
of detecting a star in such stages is very low. Our
tests showed that, as a rule, uncertainty in the assigned
ages for stars near turning points was equal
to or somewhat less than one billion years. However,
since this effect was most important for comparatively
young stars, metal-rich stars ended up being
completely dropped from the sample, since they were
taken to be located near the evolutionary stage corresponding
to the depletion of the last percent of
hydrogen in their cores and the beginning of core
collapse (they were all assigned ages that were too
young). This aspect of the method is clearly illustrated
in the age—absolute magnitude diagram for
single stars from the catalog with age uncertainties
$\epsilon t < \pm 3$ billion years, shown in Fig.~1: the complete
absence of stars at the very center of the diagram
is obvious. Therefore, because of the possible unevenness
in the inclusion of stars with younger ages,
we must treat certain features in the age — velocity-dispersion
relations with caution, especially in the
age range from three to five billion years. The ages
of stars with absolute magnitudes $M_{V} > 4.3^{m}$ are
also uncertain, since they fall in a region on the
Hertzsprung–Russell diagram where the isochrones
are very dense, making the uncertainty in their ages
very large. Fortunately, limiting the sample of stars
based on uncertainty in the ages enables us to work
with a subsample with reliable ages. Analysis shows
that, in the initial catalog, a mean uncertainty 
$\epsilon t < \pm 3$  billion years is reached for 
more than 90\,\% of the stars, and an uncertainty of 
$\epsilon t < \pm 2$ billion years for
85\,\% of the stars. Therefore, we decided that it was
reasonable to use the metallicities and ages from the
catalog~[2] for our statistical studies, bearing in mind
the problematic range of ages.

When developing a criterion for dividing up stars
according to their subsystems in the Galaxy, we took
into account their chemical compositions using data
from our master catalog of homogeneous spectroscopic
determinations of iron and magnesium abundances~[22]. 
Collected in this catalog are virtually
all magnesium (a representative of the $\alpha$\,elements)
and iron abundances in dwarfs and subgiants of the
solar neighborhood determined via synthetic fitting of
high-dispersion spectra and published up to January~2004.

\section*{COMPARISON OF A REPRESENTATIVE SAMPLE OF THIN-DISK STARS.}

The parameters of the velocity ellipsoids, by definition,
have meaning only for nearby field stars. For
stars located near the Galactic plane, the principles
used to compose a particular sample of thin-disk
stars could appreciably influence the resulting age
dependences of the axes and other parameters of the
velocity ellipsoids. It is most logical to compose a
sample based on kinematic criteria, since precisely
the velocities of the stars define the general structure
of a subsystem. However, in this case, the velocity
ellipsoids for the oldest (and consequently highest-velocity)
stellar groups will depend on some criterion
used to distinguish between the thin-disk and
thick-disk subsystems, which will be subjective to
some extent. For this reason, the index $\gamma$ in the age
dependences of the velocity dispersions, $\sigma_{i}=t^{\gamma}$, for
the thin disk could be distorted. An alternative is
to use a chemical criterion to distinguish thin-disk
stars. Many studies have indicated the existence of a
jump-like difference in the relative abundances of á
elements in thin-disk and thick-disk stars. The absolute
majority of thin-disk stars have low ratios [$\alpha/Fe]
< 0.25$ (see, for example,~[23]). Application of this
criterion enables the composition of a sample of thindisk
stars containing no more than 10\,\% of stars associated
with the kinematic thick disk; roughly the same
percent of stars that are associated with the kinematic
thin disk will not be included in the sample due to their
high relative magnesium abundances~[24]. The membership
of this small number of stars in a particular
subsystem is considered uncertain~[23].

Unfortunately, detailed atmospheric chemical
compositions require spectra of bright objects, and
so are available for only a limited number of stars. On
the other hand, since the total heavy-element content
(metallicity) can be derived from photometric data for
much fainter stars, this quantity is known for most
nearby F-G stars, making it possible to obtain statistically
significant results for these stars. The metallicity
can also be used as an independent criterion for
distinguishing thin-disk stars. The reason is that, as
was first shown in~[25], the metallicity distribution for
field stars displays a gap near $[Fe/H]\approx - 0.5$, which
naturally divides thin-disk stars from the less metalrich
stars of older subsystems of the Galaxy. As with
the relative content of $\alpha$\,elements, when this criterion
is used to select thin-disk stars, just over 10\,\% of
stars associated with the kinematic thin disk are not
included in the sample. Simultaneously, roughly this
same fraction of thick-disk stars display high enough
metallicities to be included~[24].
Thin-disk stars can also be identified based on
their ages, since this is a young subsystem, in which
mass star formation began only $\approx9$ billion years ago,
while star formation in the thick disk mostly ended
$\approx12$ billion years ago~[23]. However, due to the large
uncertainties in the ages of individual stars, we use
these ages here only in a statistical sense; i.\,e., purely
to separate stars into groups with different ages.
The above single-parameter criteria essentially
select for the same stars, and the resulting samples
of thin-disk stars differ only in stars that lie at the
border between the two disk subsystems. In~[26],
we developed a kinematic criterion for distinguishing
thin-disk stars, which implicitly takes into account
their chemical composition. This criterion is based
on our master catalog of spectroscopic derivations
of the iron and magnesium contents of 867 nearby
stars with accurate parallaxes~[26]. We distinguish
thin-disk stars simultaneouly using two kinematic
conditions, since either one of these on its own is
insufficient. The following criterion is convenient for
nearby stars: $V_{res} = (U^{2} + V^{2} +W^{2})^{1/2} < 85$ ~km\,s$^{-1}$,
where U, V , W, and $V_{res}$ are the spatial velocity
components and the residual velocity of the star
relative to the local centroid. The other criterion does
not depend on the location of a star in its Galactic
orbit, and is based on an expression proposed in~[27]:
$(Z^{2}_{max} + 4e^{2})^{1/2} < 1.05$, where $Z_{max}$ is the maximum
distance of the orbit from the Galactic plane in kpc
and e is the eccentricity of the Galactic orbit. We
assigned higher-velocity stars to the thick disk. We
chose the limiting values of both criteria so as to
minimize the number of stars with high/low relative
magnesium contents in the thin/thick disk. Unfortunately,
this criterion proved to be very strict, and
some stars with chemical compositions more suitable
for the thin disk are assigned to the thick disk.

The kinematic criterion proposed in~[28] appears to
be better justified from a methodological point of view.
In that case, the selection of thin-disk and thick-disk
stars is based on the assumption that the distributions 
of the stellar velocity components in both subsystems
relative to the local centroid are Gaussian:

\[
f(U,V,W)=\frac{1}{(2\pi)^{3/2}\sigma_{u}\sigma_{v}\sigma_{w}}
\cdot \exp(\frac{U^{2}_{LSR}}{2\sigma^{2}_{U}}-\frac{(V_{LSR}-V_{\textrm{âð}})^{2}}
{2\sigma^{2}_{V}}-\frac{W^{2}_{LSR}}{2\sigma^{2}_{W}}).
\]
Here, the ratio of the probabilities that a given star is
a member of the thin disk (D) and thick disk (TD) is
given by

\[
\textrm{D}/{\textrm{TD}}=\frac{X_{\textrm{D}}}{X_{\textrm{TD}}}
\cdot\frac{f_{\textrm{D}}}{f_{\textrm{TD}}}.
\]

Further, stars with probabilities of belonging to each
subsystem a factor of ten higher than the probability
of belonging to the other subsystem are selected. The
probabilities are determined using the known dispersions
of each of the three spatial velocity components
$\sigma_{U}$, $\sigma_{V}$, $\sigma_{W}$ the mean rotational speed  
($V_{rot}$) and the
relative numbers of stars ($X_{D}$,$X_{TD}$) in both subsystems
at the solar Galactocentric distance. The main
disadvantages of this criterion are, first, the assumed
kinematic parameters, which were derived using old
data; second, the lack of self-consistency with the
chemical composition of the stars; and, third, the fact
that stars on the boundary between the two disk subsystems
will be unidentified. As a result, this criterion
proves to be even more strict than the one described
above. If we reject the restrictions adopted in~[28]
and simply include in the thin-disk sample all stars
with a higher probability of belonging to the thin disk
than the thick disk, this sample includes stars whose
total residual velocities reach $\sim 120$~km\,s$^{-1}$; i.\,e., with
kinematic properties more typical of the thick disk.
In order to optimize this methodologically based
kinematic criterion so that it, like our criterion described
above, was simultaneously in the best possible
agreement with the elemental abundances in the
stars, we decided to refine the input parameters for
the subsystems using the data from the catalog used.
After excluding binary stars from the input catalog,
we used the method of~[28] to construct samples of
thin-disk and thick-disk stars, then calculated the
velocity-component dispersions and mean rotational
velocities for each subsystem and refined the relative
numbers of stars in both subsystems in the solar
neighborhood. This yielded for the thin disk 
$\sigma_{U} = 31$~km\,s$^{-1}$, $\sigma_{V} = 18$~km\,s$^{-1}$, 
$\sigma_{W} = 14$~km\,s$^{-1}$, and a speed
for the asymmetric shift relative to the Sun in the direction
of rotation $V_{rot} = -16$~km\,s$^{-1}$, with these values
for the thick disk being 68, 48, 49, and -53~km\,s$^{-1}$,
respectively; the relative numbers of stars in these
subsystems remained the same: 94\,\% and 6\,\%, respectively.
The largest corrections were required for $\sigma_{V}$ ,
$\sigma_{W}$, and ($V_{rot}$) for the thick-disk subsystem, which
increased by 15\,\%-20\,\%. Similar values were obtained
in~[29] (see also~[26]). We used precisely these refined
parameters to select thin-disk stars from the input
catalog for our study. No subsequent iterations were
required, since the resulting values remained virtually
constant. Figure~2a shows a plot of 
$V_{res}vs.(Z^{2}_{max} + 4e^{2})^{1/2}$, 
with the thin-disk stars selected from~[2]
using the two criteria shown with different symbols.
When the refined parameters are used, the resulting
criterion becomes close to the criterion developed
by us in~[26]. It was not necessary to include in a
given subsystem stars whose membership probabilities
were a factor of ten higher than for the other
subsystem—we simply selected stars whose membership
probability was higher for the thin than for
the thick disk. Therefore, we were able to investigate
the transition between the two subsystems, obtaining
undistorted results for samples not based on kinematic
selection criteria. Some additional stars that
were ascribed earlier to the thick disk were assigned
to the thin disk. Thirty-nine such stars were found
in our catalog of spectroscopic iron and magnesium
abundances. Tests showed that most of these stars
display low values $[Mg/Fe] < 0.25$ and high values
$[Fe/H] > -0.5$, as should be observed for thin-disk
stars. As a result, the kinematic criterion came into
better agreement with the chemical compositions.
Moreover, the resulting division of stars between the
disk subsystems appears more harmonious in the
diagram; therefore, we used this approach as a basis
for the current study.

For our analysis, we first removed known binary
stars, evolved stars ($\delta M_{V} > 3^{m}$), and stars with uncertain
ages ($\epsilon t > \pm 3$ billion years) from the input
catalog. Moreover, in order to eliminate boundary
effects, we restricted our study to effective temperatures
of $5200 - 7000$~K. After applying the criterion
developed above, 5965 nominally single F–G thin-disk
stars with reliable ages remained in the sample.
(The resulting mean uncertainty in the ages of the
sample stars was $\langle t \rangle = 1.0$ billion years.)

Since our goal here is to investigate the velocity
ellipsoids for thin-disk field stars and not stars in
moving groups, we decided to exclude only stars of
the most numerous and well known stellar flows,
without applying a cumbersome mathematical procedure
for this purpose. Guided by the mean velocity
components of these groups determined in various
studies using wavelet analyses and adaptive smoothing
(see, in particular,~[30, 31]), and also by the distribution
of our sample of thin-disk stars in the phase
space for the velocity components, we determined
approximate boundaries where the density of stars
appreciably exceeded the mean density in the field.
The Table presents the names of the stellar flows,
the ranges of velocity components relative to the Sun
used to distinguish stars in the flows, and the number
of stars excluded from the sample. The $U$, $V$ , and $W$
components are oriented toward the Galactic center,
in the direction of the Galactic rotation, and in the
direction of the North Galactic pole, respectively. The
Pleiades Group was most conveniently distinguished
using two velocity ranges. As a result, we eliminated
about 25\,\% of the sample, leaving 4557~stars.
The constructed $UV$ , $UW$ and $VW$ distributions of
the remaining sample stars demonstrate a smoother
structure. Stars belonging to other known flows that
were not removed from the sample are so few in
number in the solar neighborhood that they are not
able to significantly influence the derived parameters
of the velocity ellipsoids for thin-disk field stars.

In connection with the selection of stars in the
input Geneva–Copenhagen catalog based purely on
magnitude, the need arose to estimate the influence of
selection effects due to differences in the depths of the
survey for stars with different metallicities and temperatures.
Hotter, higher-luminosity~-— i.\,e., younger
(and therefore slower)—main-sequence stars will
dominate at large distances, distorting the velocity
ellipsoids. Figure ~2b presents the distribution of
distances from the Sun for single stars from the
catalog in the specified temperature range. The solid
curve shows a least-squares fit to the left wing of
the histogram using a power law of the form $N = \alpha R^{n}$. 
When approximating the distribution out to
60\,pc, the power-law index and its $3\sigma$ errors were
$n = 1.88 \pm 0.05$, as should be observed for a uniform
distribution of stars in the studied volume. Therefore,
for comparison, we also composed a sample out to a
distance of 60\,pc, which can be considered essentially
complete. Only 1160 thin-disk stars remained in this
sample (after excluding stars in stellar flows), which
is obviously insufficient for carrying out studies for the
entire volume; therefore, we will use this sample only
for comparison purposes.

\section*{COMPARISON OF THE VELOSITY ELLIPSOID FOR VARIOUS SAMPLES 
OF THIN-DISK STARS}

We calculated the parameters of the velocity ellipsoids
and the velocity of the Sun relative to the local
centroids using the formulas of Ogorodnikov~[32],
and calculated the corresponding errors using the
formulas of Parenago~[33]. We separated the sample
into~21 age subgroups with equal numbers of stars,
in order to trace variations of the velocity ellipsoids
with age and clarify whether the different subsamples
displayed different behavior. The first row of plots in
Fig.~3 presents the age dependences of the axes $\sigma_{i}(t)$
for correctly selected thin-disk stars, for disk stars
excluding those in moving groups, and for disk stars
located within 60\,pc of the Sun. (The corresponding
dependences for all nearby field stars are not presented
since these can be viewed in various other
studies, and, in addition, the very large scatter in the
points in the diagrams necessitates the use of different
axis scales, hindering direct comparisons; therefore,
the necessary information about these dependences
will be given in the text.) All the dependences were
approximated using least-squares power laws of the
form $\sigma_{i}\sim t^{\gamma}$.

\begin{table}
\centering
\caption{%
   Velosity components for moving-group stars relative to the Sun}
\begin{tabular}{|l|c|c|c|c|}
\hline
\multicolumn{1}{|c|}{\bf Name of group} & 
\multicolumn{1}{c|}{\bf N} &
\multicolumn{3}{c|}{\bf Velosity range of stellar streams,~km\,s$^{-1}$ }\\
\cline{3-5}
& & {\bf U} & {\bf V} & {\bf W}\\
\hline

      $\textrm{Sirius}$ & 367 &$-12\div 17 $&$ 0\div 12$  &$-18\div 6$ \\
      $\textrm{Hyades}$ & 555 &$-43\div -24$&$-23\div -10$&$-17\div -10$\\
    $\textrm{Pleiades}$ & 183 &$-17\div -5$ &$-31\div -21$&$-16\div 4$ \\
                        & 88  &$-24\div -10$&$-9\div 2$   &$-18\div 6$  \\
$\textrm{Coma Berenices}$ & 176 &$-20\div -6$ &$-10\div -2$ &$-18\div 3$\\
$\xi~\textrm{Herecules} $ & 39  &$-34\div -27$&$-52\div -44$&$-18\div 4$\\
   
\hline
\end{tabular}
\end{table}

We can estimate how much the behavior for the
corresponding axes for stars belonging to the thin
disk according to our criterion differs from that for
all nearby stars (more precisely, for all single stars of
the catalog~[2] with age uncertainties $\epsilon t < \pm3$ billion
years). The increase in the ellipsoid axes with age
can be more reliably fit with a power law for the
selected thin-disk stars than for all nearby stars.the
uncertainties in $\gamma$ for the corresponding regression
coefficients were about a factor of two smaller for all
dependences for the thin-disk stars (see for comparison
the age dependence of the velocity component
dispersions for nearby stars in Fig.~34 of~[2] and Fig.~7 of~[17]). 
At the same time, the values of the power-law
indices $\gamma$ for the same dispersions were about a
factor of 1.5~smaller. While the resulting indices for
all nearby stars are in good agreement with the typical
values obtained in other studies, $\langle\gamma\rangle=0.42$ 
on average, with uncertainties $\pm0.04$ (compare with $-0.43\pm0.04$ 
in [2,\,17]), the indices for the correctly selected
thin-disk stars were all within $3\sigma$ 
of $\langle\gamma\rangle=0.28$, with
uncertainties $\pm0.02$ (Fig.~3a). This decrease in the
power-law indices is mainly due to a decrease in
the values of all the axes for old subgroups, where
the fraction of thick-disk stars is high. Additional
analysis showed that applying stricter criteria leads to
only a slight decrease in the major axes for the oldest
subgroups, while the middle and minor axes remain
virtually unchanged. The power-law indices for all
the dependences likewise remain unchanged within
the errors. (Note that, if the sample is not restricted
to stars with age uncertainties $\epsilon t < \pm3$ billion years,
all the power-law indices grow, on average, by $\Delta\gamma\approx 0.03$.

Figure~3\,b presents the age dependences of the
axes for thin-disk field stars without stars in moving
groups. There are 217~stars in the subgroups, as
opposed to~284 in the previous sample, which led to
some increase in the uncertainties in the ellipsoid axes
(see error bars in the plots); however, the scatter of the
points was not increased, so that the uncertainty in
the power-law indices remained the same. This figure
shows that excluding stars in stellar flows increases
the ellipsoid axes for the youngest subgroups beyond
the errors, while leaving the axes for the old subgroups
virtually unchanged. As a result, the power-law
indices for all the dependences decrease somewhat,
although this change remains within the errors.

Finally, Figure~3c shows the same dependences
for thin-disk stars not in moving groups and located
within 60\,pc of the Sun. The number of stars in the
subgroups is reduced to~55, leading to a large increase
in the uncertainties of the axes and power-law indices
(see error bars and the errors in the regression
coefficients in the plots). There are no obvious shifts
in these plots compared to the others: the axes and
power-law indices for these dependences do not display
systematic variations.

Thus, applying a correct criterion for selecting
stars in the thin-disk subsystem appreciably decreased
the velocity dispersions for old groups of
stars, leaving these virtually unchanged for young
groups. This led to an appreciable decrease in the
power-law indices for the age—ellipsoid-axes relations
compared to the analogous dependences for
all nearby stars. Excluding stars in moving groups
likewise slightly reduced the power-law indices, but
this time due to an increase in the axes for young
subgroups. Restricting the analysis to the most
nearby stars, for which the sample of thin-disk stars
in the specified temperature range is complete, did not
significantly affect the results.

Note that, in all the axis plots, a break can be seen
near ages of 3.5~billion years, with a sharp rise for
small ages being replaced by a shallow relation for the
following 2.3~billion years. This is especially clearly
visible for the minor and (to a lesser extent) middle
axes. (A possible origin for this effect could be the
method used for the age determinations noted above.)

Let us now consider how the other parameters of
the velocity ellipsoids depend on the sample used.
The second row of plots in Fig.~3 presents the age
dependences of the ratios of the axes for the same
samples. Linear approximations of the points on all
the graphs indicate small positive trends for the ratios
of the minor and middle axes to the major axis,
although the probability that these trends arose by
chance were much higher than 5\,\%, reflecting their
small statistical significance. On all the graphs, the
ratios of the middle to the major axis were $< 0.60$ for
all ages (the larger scatter of the points about the fit
curves and the larger error bars in the last graph are
due primarily to the fact that the subgroups contain
only a quarter as many points). Both ratios display
some (though statistically insignifcant) variation in
the slopes of the age dependenes beyond ages of
$\approx 3-5$ billion years.

The third row of plots in Fig.~3 presents the age
dependences of the vertex coordinates in Galactic
coordinates for the same samples. Linear fits for all
the samples display a weak decrease in the deviations
in the vertex longitude $(L)$ with increasing age
(especially for the sample excluding moving groups),
while the vertex latitude does not vary systematically
with the age, being roughly equal to zero in all the
diagrams. All these weak correlations are formally
insignificant, although the constancy of the sign of
the slopes for the vertex deviations in longitude for the
various samples may provide evidence for their reality;
it may be that these correlations would become significant
for a larger sample. All the samples display
distortions of the monotonicity of the age dependence
of the Galactic longitude of the vertex direction near
ages of 3--5~billion years (Figs.~3g, 3h, 3i).

The age dependences of the integrated parameters
of the velocity ellipsoids for the components of the
solar motion relative to the corresponding centroids
and the Galactic coordinates of the apex of the solar
motion, presented in the fourth and fifth rows of Fig.~3,
are nearly identical for all three samples considered.
The $U$ and $W$ components of the solar velocity are
nearly independent of the age, while the $V$ component
increases significantly with age (as is reflected by the
high correlation coefficient for this relation). Since the
solar motion reflects the motion of the corresponding
centroids, this indicates that, on average, the angular
momentum of the stars decreases linearly with their
age. The high correlation coefficients for the linear fits
presented in the last row of Fig.~3 indicate a significant
increase in the longitude and decrease in the
latitude of the apex of the solar motion with increasing
age.

\section*{VELOCITY ELLIPSOIDS FOR STARS WITH VARIOUS METALLICITIES}

Analyzing the chemical composition of thin-disk
stars, we found that a larger percent with anomalous
relative magnesium overabundances appears with increasing
metallicity~[24]. This can be interpreted as
evidence that these stars were formed from material
that underwent a different chemical, and possibly
dynamical, evolution. While stars ''remember''
the dynamical state of the interstellar medium from
which they formed, the kinematics of stars of different
metallicities can be different. In this connection, it is
interesting to consider the velocity ellipsoids for stars
with different metallicities. Let us first consider how
the velocity ellipsoids for stars of mixed ages depend
on metallicity. Since restricting the distances of the
stars did not lead to any distortion of the velocity
ellipsoids, we will further use our complete sample
of thin-disk stars excluding stars in moving groups.
To construct the diagrams in Fig.~4, the sample was
divided into 11~subgroups with different metallicities,
each containing 414~stars. Most of the parameters
display trends with varying metallicity (see the corresponding
correlation coefficients on the plots). For
example, all the axes systematically decrease with
increasing metallicity, with the probability that this
came about by chance being $P \ll 1\,\%$ for all the
correlations. The ratio of the middle or minor axes to
the major axis, as well as the deviation of the vertex
latitude also decrease with increasing metallicity, although
these variations have low significance ($P \approx 20\,\%$), while the longitude deviations of the vertex
increase strongly ($P \ll 1\,\%$). Some integrated parameters
of the motion of the velocity ellipsoids relative
to the Sun also display systematic variations with
metallicity. Figure~4 shows that the correlations for
the azimuthal velocity component and for both coordinates
of the apex are highly significant ($P \ll 1\,\%$).
The two other components of the solar motion are
independent of metallicity. The $V$ component displays
a paradoxical behavior (first noted in~[34]): the angular
momentum of more metal-rich groups is lower than
that of less metal-rich groups (and not higher, as
could be expected!). Indeed, we saw above that the
rotational velocity of stars about the Galactic center,
on average, decreases linearly with increasing age, so
that the angular momentum (which is proportional
to the rotational velocity for nearby stars) can be
taken as a statistical age indicator. On the other
hand, for equal temperatures, the ''isochrone'' (real)
ages of more metal-rich main-sequence stars are, on
average, lower than those for their less metal-rich
counterparts. As a result, for equal ''isochrone'' ages,
more metal-rich stars are generally ''kinematically''
older than less metal-rich stars.

One feature that appears to some extent in all
of the studied metallicity dependences is a break
near $[Fe/H] \approx -0.25$. This break can be traced most
clearly in the relations for $\sigma_{2}/\sigma_{1}$, the vertex deviations
(L), and the longitude of the apex; the correlation
essentially disappears to the right of the break. A
''leveling off'' of the curves to the right of the break can
also be seen in the dependences for all the ellipsoid
axes.

Let us consider further the age dependences for
stars in narrow metallicity ranges. For this, we divided
the sample of thin-disk stars excluding stars
in moving groups into four roughly equal groups
separated by [Fe/H] values of $-0.25$, $-0.13$, and
0.00, then divided each into 12 subgroups according
to age. The plots in the first row of Fig.~5 show
the age dependences of the axes and dispersions of
the total residual velocities for the various metallicity
groups. The power-law character of the dependences
is maintained in each metallicity group. Systematic
variations in the power-law indices with metallicity
can be traced for all the axes and the dispersion of
the total residual velocity, in spite of the relatively
low number of stars in the subgroups (approximately
95), which led to errors much larger than those in
Fig.~3. All the power-law indices for a given dependence
display the largest values for the group with intermediate
metallicity ($-0.13 < [Fe/H] < 0.00$), and
systematically decrease on either side of this group
(an exception is the value of $\gamma$ for the minor axis of the
lowest-metallicity group). The decrease in the power-law
indices for the metal-poor stars is due to the
increase in the velocity dispersions for the youngest
subgroups, and the decrease for metal-rich stars to
the decrease in the velocity dispersions for old groups.
The firm detection of systematic differences in the
power-law indices for stars with different metallicities
will require a larger sample of thin-disk stars. A comparison
of the plots in the first row also indicates that
the systematic increase in the velocity dispersions
with increasing metallicity for stars of mixed ages
visible in Fig.~4a can be traced in stars of any age.
The lower the metallicity of the group, the higher, on
average, the dependences of a given type lie, with this
rise increasing with age. Note that signs of the break
near 3.5 billion years detected in the dependences
for all thin-disk stars (Fig.~3) can be traced in all the
groups.
The plots in the middle row of Fig.~5 present
the age dependences for the axis ratios for the same
groups. These slopes do not display systematic variations
with metallicity, although all the age - $\sigma_{i}/\sigma_{j}$
correlations are positive, with the correlation coefficients
differing from zero beyond the errors in some
cases. More reliable determination of the slopes of
these relations and differences between them will also
require a larger sample of thin-disk stars.
The age and metallicity dependences for the directions
of the vertices of the velocity ellipsoids are
presented in the third row of plots in Fig.~5. In all the
plots, the deviations of the vertices in neither longitude
nor latitude display significant age dependences:
all the correlation coefficients are equal to zero within
the errors. The complete absence of systematic
variations is obvious: the corresponding correlations
sometimes change sign in the transition between
metallicity groups. At the same time, the systematic
increase in the vertex longitude with increasing
metallicity for mixed-age stars visible in Fig.~4c is
characteristic for stars of all ages. Thus, the major
axes of the velocity ellipsoids for stars of any age
are oriented virtually toward the Galactic center for
stars of low metallicity, but deviate more and more
from this direction with increasing [Fe/H], but only
in longitude.

Let us now turn to parameters that characterize
the motion of the groups as a whole rather than the
dynamical state of the stars within a group. The first
row of plots in Fig.~6 presents the age and metallicity
dependences for the velocity components of the
solar motion relative to the local centroids for the
corresponding stellar groups. The negatives of these
velocities represent the motion of the local centroids
relative to the Sun. The components of the solar motion
toward the Galactic center display a large scatter
about the linear lines fit in all of the diagrams, and
are independent of age (see corresponding correlation
coefficients). The highly significant variations of V
($P \ll 1\,\%$), characterizing the angular momentum,
indicate that the rotational speeds about the Galactic
center for groups of any metallicity decrease with increasing
age. The slopes of these dependences are approximately
the same within the errors, and the paradoxical
decrease in angular momentum with increasing
metallicity is observed for all ages—the $V_{\odot} -t$
relation simply shifts upward with increasing metallicity.
The velocity components directed perpendicular
to the Galactic plane are roughly the same for all the
groups, and do not depend on age. These properties of
the velocity components are also reflected in the behavior
of the coordinates of the apex of the solar motion
relative to the stellar groups with different ages
and metallicities—significant trends demonstrating
an increase in Galactic longitude and decrease in
Galactic latitude with age are clearly visible (second
row of plots in Fig.~6). The age dependence of the
apexlon gitude is strongest for the lowest-metallicity
group, but then weakens abruptly, purely due to the
increase in the apexlongitu de relative to young stellar
subgroups; further, in the range $[Fe/H] > -0.25$, no
systematic variations with metallicity are seen for
either the slopes or the positions of the dependences.
In the transition to more metal-rich groups, the shifts
in latitude display a gradual decrease in their slopes
in the entire metallicity range characteristic for the
Galactic disk, likewise due exclusively to the decrease
in the latitude of the solar-motion apex relative to
young stars.

\section*{DISCUSSION}

Holmberg et\,al.~[2] constructed age - velocity-component
dispersion relations for nearby stars with
reliably determined ages and approximated them
using power laws. The resulting power-law indices
$\gamma$ were 0.38, 0.38, 0.54, and 0.40 for $\sigma_{U}$, 
$\sigma_{V}$, $\sigma_{W}$,
and $\sigma_{tot}$, respectively (or 0.39, 0.40, 0.53, and 0.40
in the study of Holmberg et\,al.~[17]). Here, we have
calculated the velocity-ellipsoid semi-axes, which,
in general, can differ slightly from the velocity-component
dispersions in a Galactic orthogonal
coordinate system, since, as a rule, the major axis
of the ellipsoid does not coincide with the direction
toward the Galactic center. As a result, the indices for
the age.axis relations for nearby stars obtained using
the same data are slightly different: 0.38, 0.39, 0.48,
and 0.42 for $\sigma_{1}$, $\sigma_{2}$, $\sigma_{3}$ and 
$\sigma_{tot}$, respectively. After
excluding thick-disk stars from the sample, all the
power-law indices decreased appreciably (Fig.~3a).
They decreased overall by approximately a factor
of $1.5 - 2$ after further excluding known members of
stellar flows, becoming 0.22, 0.26, 0.27, and 0.24
(Fig.~3b). The index for the velocity component
perpendicular to the Galactic plane decreased most
of all, although, as before, it remained the largest.

The age dependence of the W velocity component
for nearby stars was considered in some detail
in~[16], using data from the catalog~[15]. The main
result of that study was to refute the results of [15,\,17] 
concerning the continuous growth of the dispersion
of W with age in the thin disk, instead finding
evidence that this growth saturated $\approx4.5$ billion
years ago. As a consequence, the age - $\sigma_{W}$ relation
should be fit using two power laws that intersect near
4.5~billion years. Seabroke and Gilmor~[16] propose
a fundamentally new mechanism for this behavior.
interaction with another galaxy. Indeed, we can see a
gleveling offh of the age~-- $\sigma_{3}$ relation near 3.5~billion
years in Fig.~3c (with a similar ''leveling off'' also
being appreciable for the other two axes). However,
as we noted, this may be an artefact associated with
the method used to determine the most probable ages
of the stars in the input catalog. Therefore, this result
must be verified using independent stellar ages
derived using other methods.

Thus, we have shown that the velocity ellipsoid
for thin-disk stars increases in size with increasing
age, while its shape remains nearly constant (or only
slightly increases). The direction of the major axis
also remains nearly constant with age.the ellipsoid
retains a constant, non-zero vertex deviation. At the
same time, the ellipsoid's motion in the Galaxy is
such that its angular momentum decreases appreciably
with increasing age, while the two other velocity
components remain constant. In a stationary rotating
stellar system, the condition 
$\sigma_{2}/\sigma_{1} = [-B/(A-B)]^{1/2}$ 
should be satisfied. If the Oort constants $A$ and
$B$ are taken to be 13.7 and -12.9~km\,s$^{-1}$~kpc [35],
the ratio of the middle to the major axis in the Galaxy
should be~0.70. Since we have found this ratio to
be smaller for all ages, 
$(\langle\sigma_{2}/\sigma_{1}\rangle = 0.58 \pm 0.03)$, this
suggests that the Galaxy should be considered to be
far from stationary at any time. (Although this last
conclusion can be avoided by adopting values for the
Oort constants, for which the axis ratio should be
equal to 0.58 (15 and -10 ~km\,s$^{-1}$~kpc)).
The parameters of the velocity ellipsoids depend on
metallicity. With decreasing metallicity, the velocity
ellipsoids increase in size, become more spherical,
and turn in longitude toward the Galactic center. We
have found an obviously paradoxical behavior for the
angular momentum: for stars of equal ages, the rotational
velocity about the Galactic center decreases
with increasing metallicity. As a result, the mean
angular momentum of the most metal-rich disk stars
with ages of $\approx 2$ billion years is the same as that of the
least metal-rich stars with ages of  $\approx 8$ billion years.
This cannot be explained by relaxation processes in
the Galactic disk, since these cannot lead to systematic
variations in the velocities of stars of a given
metallicity.

\section*{CONCLUSION}
Let us say a few words about the origin of kinematic
differences among stellar populations in the
Galactic thin disk. Our results testify that relaxation
processes due to spiral density waves are able to
explain the power-law character of the increase in
the dispersions of the U and V stellar-velocity components
in the directions toward the Galactic center
and of the Galactic rotation with increasing age,
with the power-law indices being $\gamma_{1,2} \approx 0.25$, and not
0.3--0.4, as has usually been thought. The somewhat
higher index we have obtained for the dispersion of
the vertical velocity component, $\gamma_{3} \approx 0.27$, coincides
within the errors with the theoretical upper limit associated
with ''heating'' of stars by giant molecular
clouds alone, without the participation of massive
objects residing in them. If we suppose that relaxation
processes play a determining role in the evolution
of the disk, it remains unclear why the shape of the
velocity ellipsoid should display only a very weak time
dependence, while simultaneously remaining far from
equilibrium. On the other hand, the significant decrease
in the mean angular momentum of stars with
increasing age testifies that the stars ''remember''
the dynamical state of the interstellar medium from
which they formed, and that relaxation processes have
not fully distorted the initial stellar-velocity distribution.
The existence of metallicity dependences for the
velocity ellipsoids for stars of the same age is also
difficult to explain purely using relaxation processes,
although corresponding age dependences for different
metallicity groups differ little. The larger radial and
azimuthal velocity dispersions and nearly zero vertexdeviations
of stars with $[Fe/H] < -0.25$ (especially
young ones) provide evidence that they formed
from interstellar gas containing a substantial fraction
of matter that had fallen in from outer parts of the
Galaxy. Support for this hypothesis comes from their
enhanced abundances of $\alpha$\,elements, compared to
the Sun~[24]. However, the high angular momentum
carried by young, metal-poor stars and their small (as
for other young disk stars) velocity dispersions perpendicular
to the Galactic disk are not fully in agreement
with this picture. It is also difficult to understand
the linear growth in the vertex deviation, which is
the same for stars of any age, and the paradoxical
decrease in angular momentum with increasing
metallicity. Another problem that may be related to
these is the sharp decrease in the power-law indices
for the age dependences of the velocity dispersions for
stars with $[Fe/H] > 0.0$. One of the most promising
ways to explain all these contradictions may be a detailed
analysis of radial migrations of stars of various
metallicities, which have ended up near the Sun at
the current time as a result of the action of inhomogeneities
in the Galactic potential. The analysis of
the orbits of nearby stars carried out in~[36] showed
that more than half of stars with $[Fe/H] < -0.4$ have
orbits whose properties suggest they were born in the
outer regions of the Galactic disk, while most stars
with $[Fe/H] > 0.1$ were born in inner regions.

Thus, our use of the largest sample of nearby
field stars with accurately determined physical and
kinematic parameters currently available has enabled
us to identify several important dependences, that
can ultimately provide a better understanding of the
evolution of the thin disk. Nevertheless, as we have
noted in the text, we were not able to draw statistically
significant conclusions in some cases.


\newpage

\begin{figure*}
\centering
\includegraphics[angle=0,width=0.96\textwidth,clip]{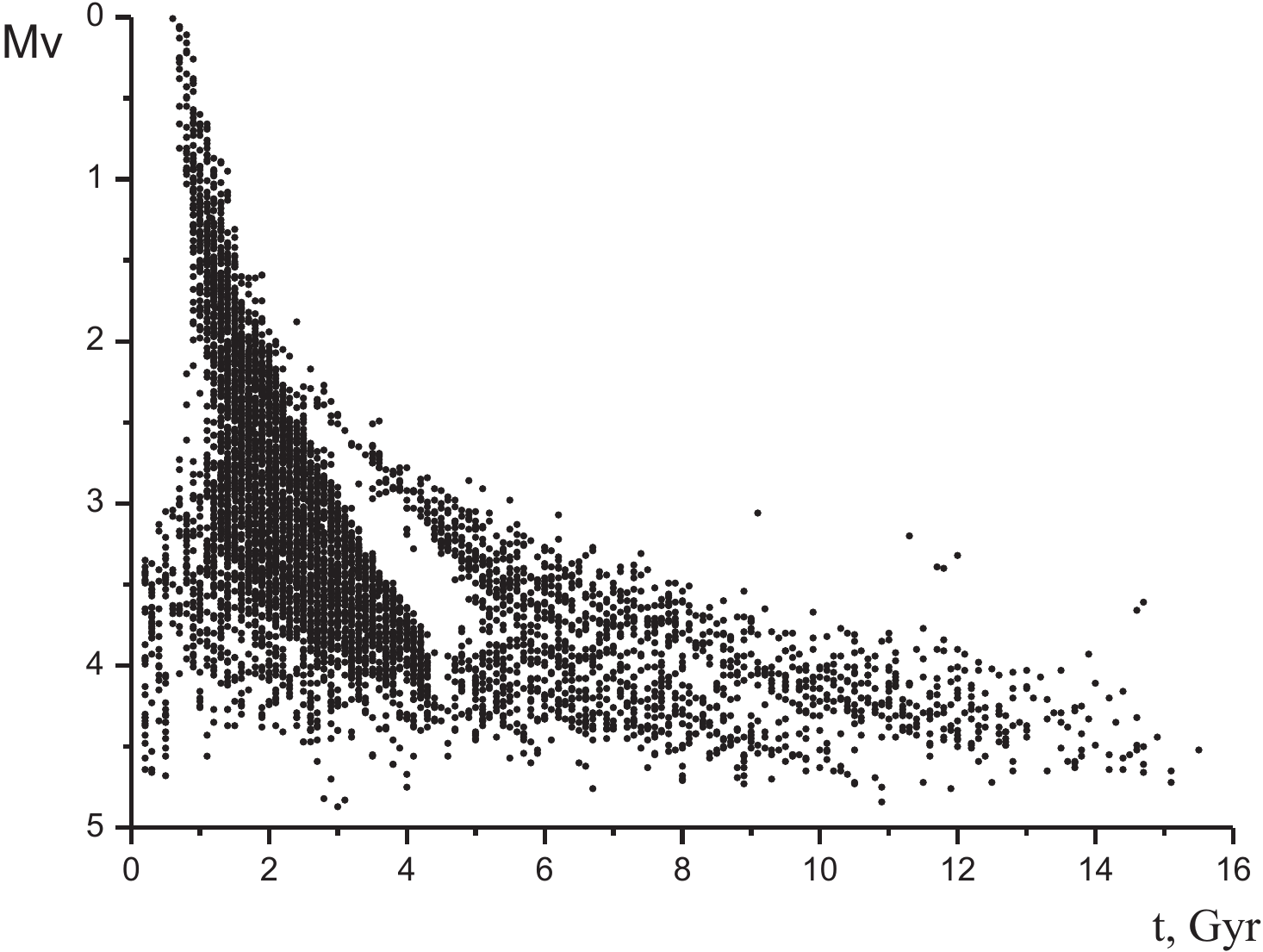}
\caption{Age—absolute magnitude diagram for single stars of the 
catalog [2] with age uncertainties $\epsilon t < \pm 3$ billion years.}
\label{fig1}
\end{figure*}

\newpage

\begin{figure*}
\centering
\includegraphics[angle=0,width=0.96\textwidth,clip]{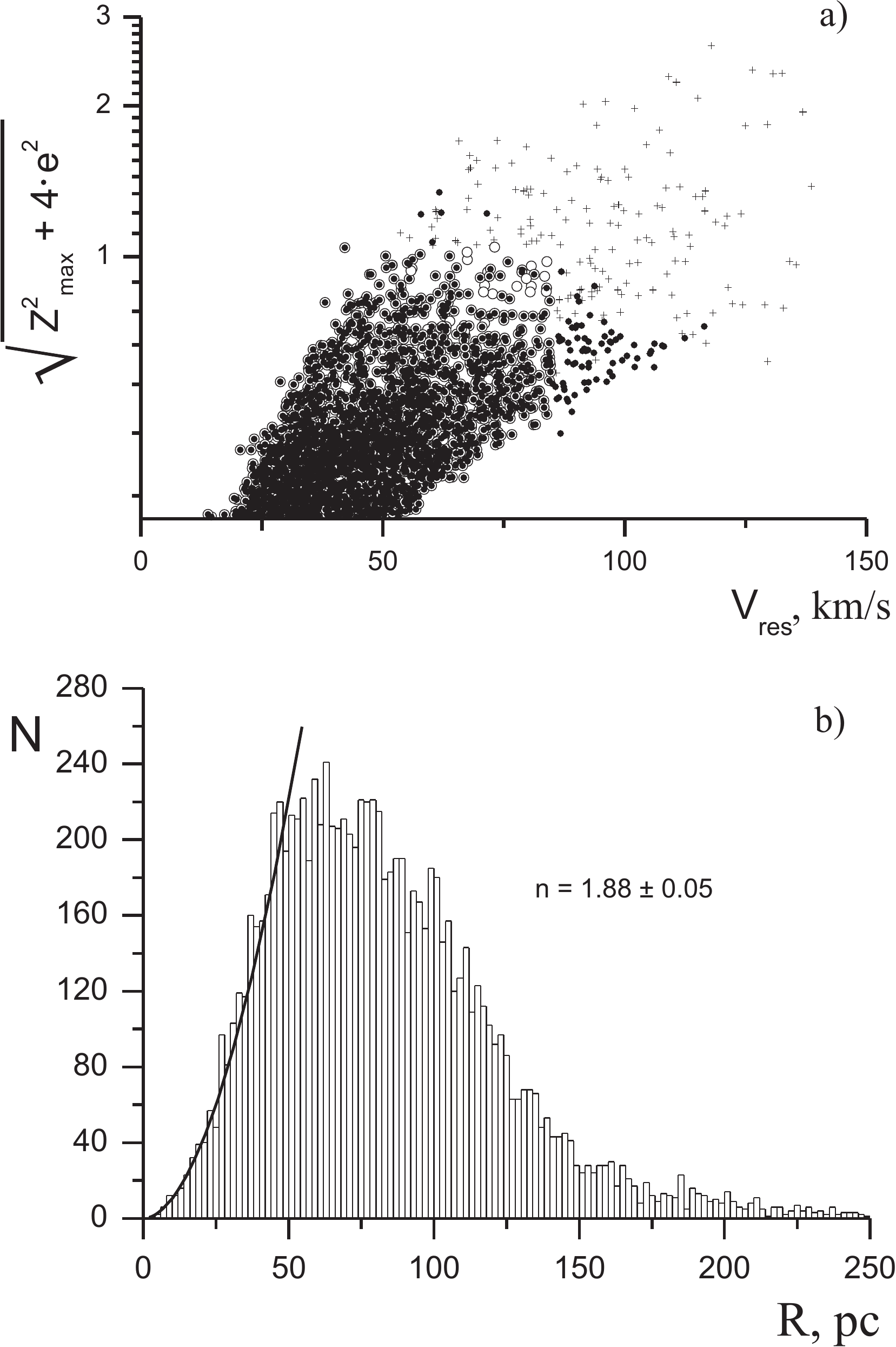}
\caption{(a) Relation between $\sqrt{Z^{2}_{max} + 4e^{2}}$ and 
the residual velocities relative to the local centroid for stars 
with age uncertainties $\epsilon t < \pm 3$~Gyr and (b) the 
distribution of distances from the Sun for single stars from the 
catalog~[2]. Hollow circles show thin-disk stars selected according 
to the criterion of~[26], and filled circles stars selected according 
to the new criterion. The pluses show all remaining stars of the 
catalog. The curve is a power-law fit to the distribution within 
60\,pc from the Sun.}
\label{fig2}
\end{figure*}

\newpage

\begin{figure*}
\centering
\includegraphics[angle=0,width=0.96\textwidth,clip]{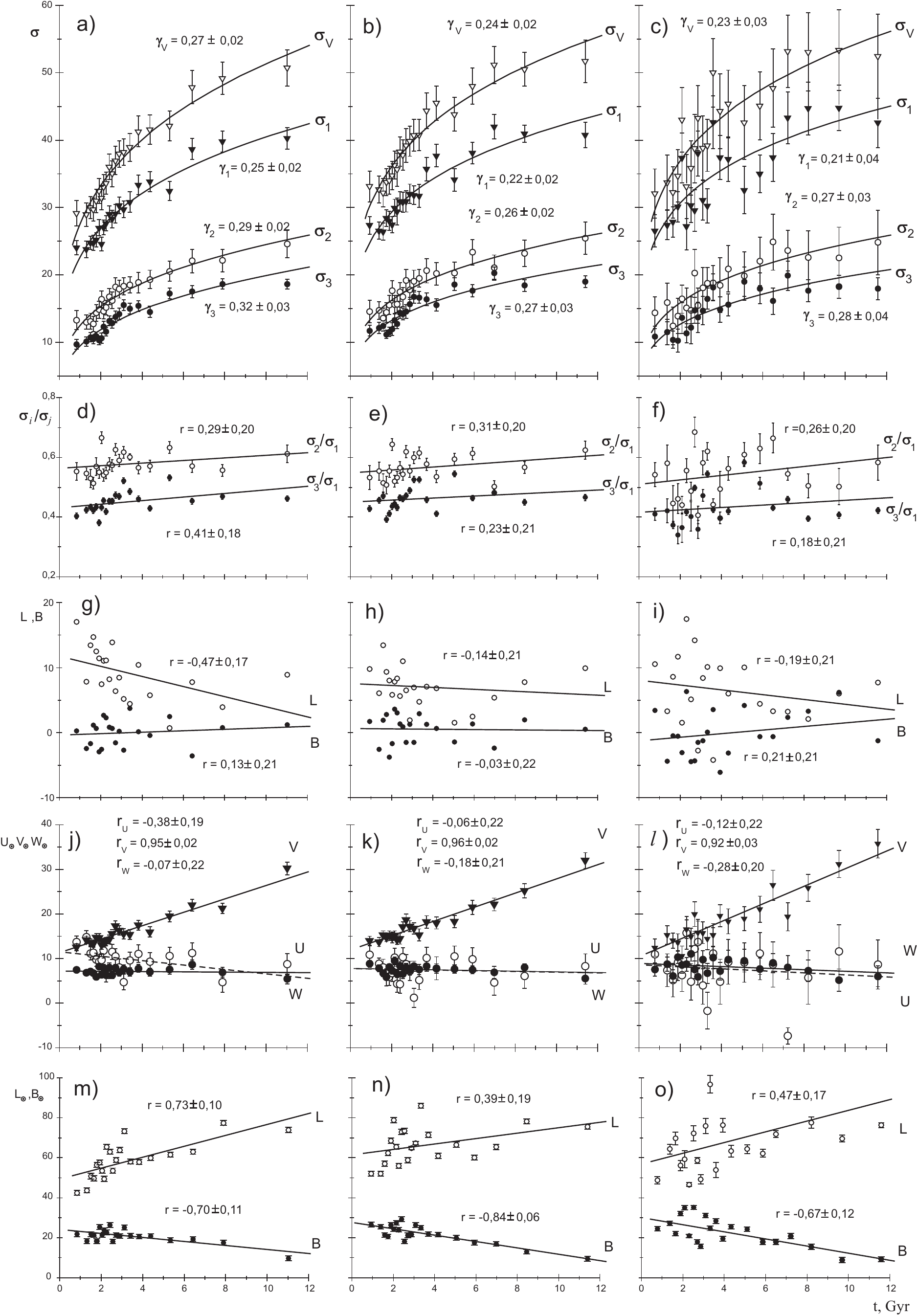}
\caption{Age dependences of the velocity ellipsoids for all single 
F-–G thin-disk stars with age uncertainties $\epsilon t < \pm 3$~Gyr. 
The first column shows plots for stars selected according to our 
criterion, the second excludes stars of moving groups,
and the third excludes stars more than 60\,pc from the Sun. The age 
dependences for the axes of the velocity ellipsoids and
the total residual velocity (upper row), ratios of the ellipsoid axes 
(second row), vertex coordinates (third row), solar-velocity
components relative to the corresponding centroids (fourth row), and 
apex coordinates for the solar motion (fifth row) are
presented. The curves in the upper row of plots are power-law fits, the 
titles above the plots give the power-law indices with
their errors, and the bars show the errors in the corresponding 
quantities (for the vertex and apex coordinates, these are smaller
than the symbols). On the remaining graphs, the solid curves show 
linear fits and the numbers indicate the corresponding
correlation coefficients.}
\label{fig3}
\end{figure*}

\newpage

\begin{figure*}
\centering
\includegraphics[angle=0,width=0.96\textwidth,clip]{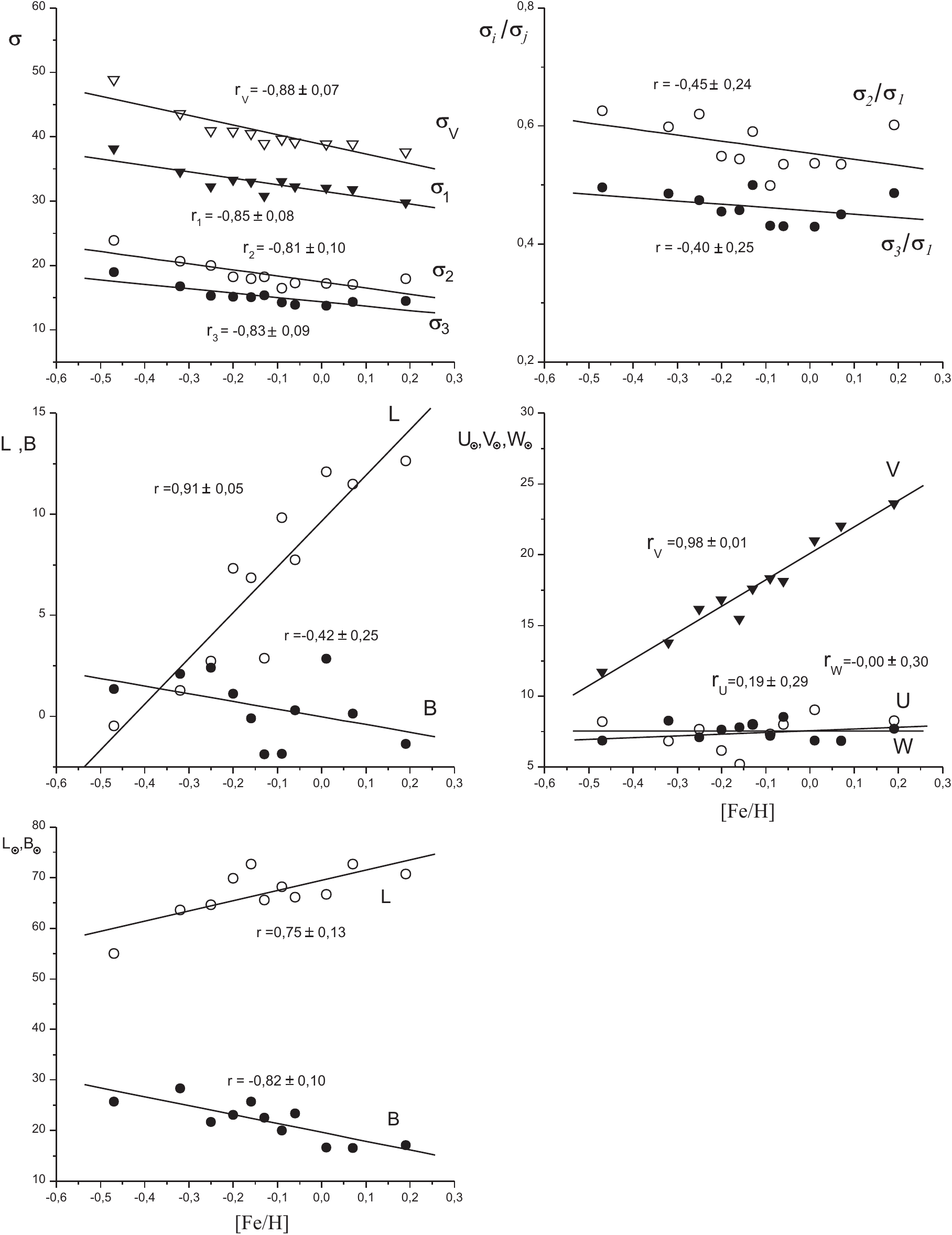}
\caption{Metallicity dependences of the (a) velocity-ellipsoid 
axes and the total residual velocity, (b) the ratios of the 
ellipsoid axes, (c) the coordinates of the vertex, (d) the 
components of the solar velocity relative to the corresponding 
centroids, and (e) the coordinates of the apex of the solar motion. 
The solid lines are linear fits and the numbers indicate the 
corresponding correlation coefficients. The mean error bars for the 
parameters are less than the size of the symbols in all the figure 
panels.}
\label{fig4}
\end{figure*}

\newpage

\begin{figure*}
\centering
\includegraphics[angle=90,width=0.96\textwidth,clip]{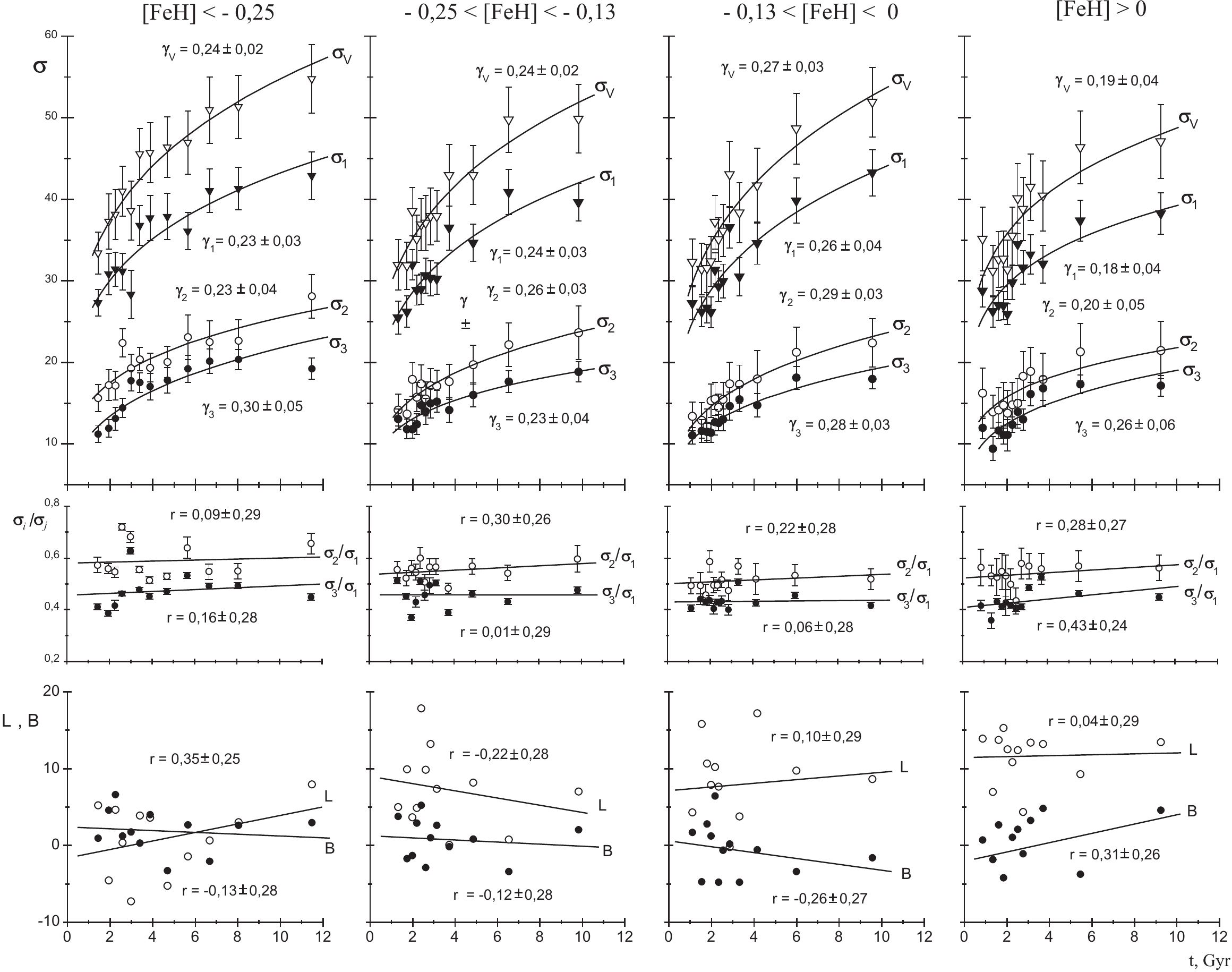}
\caption{Age dependences of the velocity ellipsoids for thin-disk 
stars for four metallicity groups. The metallicity ranges are indicated 
above. The notation is as in Fig.~3.}
\label{fig5}
\end{figure*}

\newpage

\begin{figure*}
\centering
\includegraphics[angle=90,width=0.96\textwidth,clip]{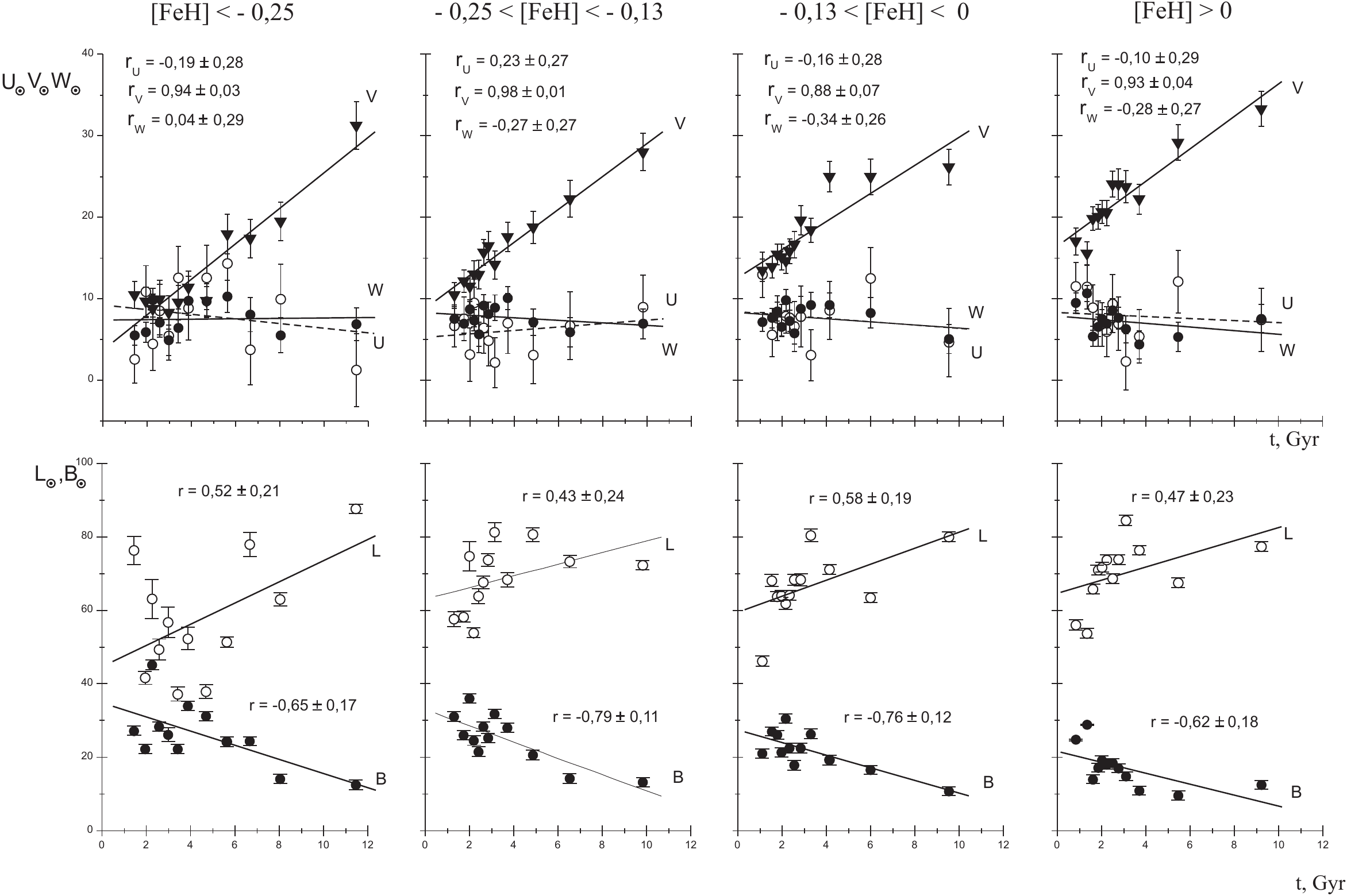}
\caption{Age dependences of the solar velocity components relative 
to the corresponding centroids (first row) and the coordinates of 
the solar-motion apex( second row) for four metallicity groups. The 
metallicity ranges are indicated above. The notation is as in Fig.~3.}
\label{fig6}
\end{figure*}

\end{document}